\font\bbf=cmbx12

\def \Ezp	{{\bf E}^{zp}}
\def \Bzp	{{\bf B}^{zp}}

\def \< {\left\langle}
\def \> {\right\rangle}
\def \[ {\left[ }
\def \] {\right] }
\def \( {\left(}
\def \) {\right)}

\hfill {\bf AIAA 2001-3360}
\vskip 0.4truein
\par\centerline{\bbf Geometrodynamics, Inertia and the Quantum Vacuum}
\vskip 0.2truein
\bigskip\centerline
{\bf Bernard Haisch}
\medskip\centerline{
California Institute for Physics and Astrophysics, 366 Cambridge Ave., Palo Alto, CA
94306}
\par\centerline{$<$www.calphysics.org$>$}
\par\centerline{Associate Fellow, AIAA --- \it haisch@calphysics.org}
\bigskip\centerline
{\bf Alfonso Rueda}
\par\centerline
{Department of Electrical Engineering, ECS Building}
\par\centerline
{California State University, 1250 Bellflower Blvd.,
Long Beach, California 90840}
\par\centerline
{\it arueda@csulb.edu}
\bigskip\noindent
\bigskip

\centerline{({\it  Presented at the AIAA/ASME/SAE/ASEE Joint Propulsion Conference, Salt Lake City,
July 8--11, 2001})}

\bigskip
\centerline{\bf ABSTRACT}
\bigskip
Why does {\bf F} equal m{\bf a} in Newton's equation of motion?
How does a gravitational field produce a force?
Why are inertial mass and gravitational mass the same?
It appears that all three of these seemingly axiomatic foundational questions have an
answer involving an identical physical process: interaction between the electromagnetic quantum
vacuum and the fundamental charged particles (quarks and electrons) constituting matter. All three
of these effects and equalities can be traced back to the appearance of a specific asymmetry in the
otherwise uniform and isotropic electromagnetic quantum vacuum. This asymmetry gives rise to a
non-zero Poynting vector from the perspective of an accelerating object. We call the resulting
energy-momentum flux the {\it Rindler flux}. The key insight is that the asymmetry in an accelerating
reference frame in flat spacetime is identical to that in a stationary reference frame (one that is
not falling) in curved spacetime. Therefore the same Rindler flux that creates inertial reaction
forces also creates weight. All of this is consistent with the conceptualizaton and formalism of
general relativity. What this view adds to physics is insight into a specific physical process
creating identical inertial and gravitational forces from which springs the weak principle of
equivalence. What this view hints at in terms of advanced propulsion technology is the possibility
that by locally modifying either the electromagnetic quantum vacuum and/or its interaction with
matter, inertial and gravitational forces could be modified.

\bigskip
\centerline{\bf INTRODUCTION}
\bigskip
\centerline{Why does {\bf F} equal m{\bf a} in Newton's equation of motion, ${\bf F}=m{\bf a}$?}
\par
\centerline{How does a gravitational field produce a force?}
\par
\centerline{Why are inertial mass and gravitational mass the same?}
\bigskip
These are questions that are usually thought to be more appropriate for philosophers than for
physicists, since the apparent facts of nature addressed in these questions are generally regarded as
axiomatic.[1] If one assumes that one plus one equals two, plus a limited number of additional axioms,
one can develop a self-consistent system of mathematics, but one has to start with such
assumptions. In the realm of geometry, the discovery of non-Euclidean geometry in the 19th century
taught us that other, non-intuitive assumptions are possible, and indeed, Riemannian geometry became
the basis for physics in Einstein's general relativity (GR). Alternate foundational assumptions can
lead to new insights.

\bigskip
Since 1991 we have been engaged in investigations involving the nature of inertia based on the
foundational assumption that the electromagnetic quantum vacuum, also called the zero-point field or
zero-point fluctuations, is a real and underlying universal sea of energy capable of interacting with
matter, and that this interaction may be described using the techniques of stochastic electrodynamics
(SED).[2] Following a successful multiyear NASA-funded investigation at Lockheed Martin and
California State University at Long Beach, the privately-funded California Institute for Physics and
Astrophysics (CIPA) was established in 1999 specifically to study the electromagnetic quantum vacuum
and its effects. A group of five postdoctoral fellows with expertise in quantum electrodynamics,
superstring and M-brane theory, general relativity, plasma physics and Casimir effects are engaged in
theoretical efforts to come to a deeper understanding of foundational questions in physics and to
examine the possible nature and degree of electromagnetic quantum vacuum--matter interactions both
within and beyond the SED approximations.
CIPA has also funded projects by experts at several universities directed toward the general goal of
explaining the origin of inertia and certain aspects of gravitation and other relevant physical and
astrophysical effects as due to the electromagnetic quantum vacuum.

\bigskip
It appears that all three of the seemingly axiomatic foundational questions posed above have an
answer involving an identical physical process: interaction between the electromagnetic quantum
vacuum and the fundamental charged particles (quarks and electrons) constituting matter. All three
of these effects and equalities can be traced back to the appearance of a specific asymmetry in the
otherwise uniform and isotropic electromagnetic quantum vacuum. The key insight is that the asymmetry
in an accelerating reference frame in flat spacetime is identical to that in a stationary reference
frame in curved spacetime.

\bigskip
It was shown by Unruh [3] and by Davies [4] that a uniformly-accelerating detector will
experience a Planckian-like heat bath whose apparent ``temperature'' is a result of quantum vacuum
radiation. A tiny fraction of the (enormous)
electromagnetic quantum vacuum energy can emerge as real radiation under the appropriate conditions.
The theoretical prediction of Unruh-Davies radiation is now generally accepted and SLAC physicist P.
Chen has recently proposed an experiment to detect and measure it.[5] Rueda and Haisch (hereafter RH)
[6] analyzed a related process and found that as perceived by an accelerating object, an energy and
momentum flux of radiation emerges from the electromagnetic quantum vacuum and that the strength of
this momentum flux is such that the radiation pressure force on the accelerating object is
proportional to acceleration. Owing to its origin in an accelerating reference frame customarily
called the {\it Rindler frame}, and to the relation of this flux to the existence of a {\it Rindler
event horizon}, we call this flux of energy and momentum that emerges out of the electromagnetic
quantum vacuum upon acceleration the {\it Rindler flux}.

\bigskip
If the Rindler flux is allowed to
electromagnetically interact with matter, mainly but perhaps not exclusively at the level of quarks
and electrons, a reaction force is produced that can be interpreted as the origin of Newton's ${\bf
F}=m{\bf a}$. In this view, which we call the {\it quantum vacuum inertia hypothesis}, matter resists
acceleration not because of some innate property of inertia, but rather because the 
quantum vacuum fields provide an acceleration-dependent drag force. In future attempts we and
coworkers intend to examine the possible contributions of other components of the quantum vacuum
besides the electromagnetic. (This is relevant to the issue of possible neutrino mass, which could not
be due to the electromagnetic quantum vacuum, but might possibly be due to the vacuum fields of the
weak interaction.)

\bigskip
GR declares that gravity can be interpreted as spacetime curvature. Wheeler
coined the term geometrodynamics to describe this: the dynamics of objects subject to gravity is
determined by the geometry of four-dimensional spacetime. What geometrodynamics actually specifies is
the family of geodesics --- the shortest four-dimensional distances between two points in spacetime 
--- in the presence of a gravitating body. Freely-falling objects and light rays follow geodesics.
However when an object is prevented from following a geodetic trajectory, a force is experienced: the
well-known force called weight. Where does this force come from? Or put another way, how does a
gravitational field exert a force on a non freely-falling, fixed, object, such as an observer
standing on a scale on the Earth's surface? {\it This proves to be the identical physical process as
described in the quantum vacuum inertia hypothesis, due to a non-zero Rindler flux.}

\bigskip
In the SED approximation, the electromagnetic quantum vacuum is represented as propagating
electromagnetic waves.[7,8] These should follow geodesics.
It can be shown that propagation along curved geodesics creates the identical electromagnetic
Rindler flux with respect to a stationary fixed object as is the case for an accelerating object.[9]
This is perfectly consistent with Einstein's fundamental assumption of the equivalence of gravitation
and acceleration. An object fixed above a gravitating body will perceive the electromagnetic quantum
vacuum to be accelerating past it, which is of course the same as the perception of the object when
it is doing the accelerating through the quantum vacuum.
Thus in the case of gravity, it would be the electromagnetic Rindler flux acting upon a fixed object
that creates the force known as weight, thereby answering the second question. The answer to the
third question then immediately follows. Since the same flux would be seen
by either a fixed object in a gravitational field or an accelerating object in free space, the force
that is felt would be the same, hence the parameters we traditionally call inertial and
gravitational mass must be the same. This would explain the physical origin of the weak principle of
equivalence.

\bigskip
All of this is consistent with the mathematics of GR. What this view adds to physics is insight into
a specific physical process creating identical inertial and gravitational forces. What this view
hints at in terms of advanced propulsion technology is the possibility that by locally modifying
either the quantum vacuum fields and/or their interaction with matter, inertial and
gravitational forces could be modified and possibly one day freely controlled.

\bigskip
\centerline{\bf THE ELECTROMAGNETIC QUANTUM VACUUM}
\bigskip
The quantization of the
electromagnetic field in terms of quantum-mechanical operators may be found in various standard
textbooks, e.g. Loudon [10]: ``The electromagnetic field is now quantized by the
association of a quantum-mechanical harmonic oscillator with each mode {\bf k} of the radiation
field." This can be understood as follows: Application of the Heisenberg uncertainty relation to a
harmonic oscillator requires that its ground state have a non-zero energy of $h\nu/2$. This reflects
the fact that quantum mechanically a particle cannot simultaneously be exactly at the bottom of its
potential well and have exactly zero momentum. There exists
the same
$h\nu/2$ zero-point energy expression for each mode of the electromagnetic field as for a mechanical
oscillator. (Formally, mode decomposition yields that each mode can be mathematically made into a
harmonic oscillator in the sense that the same differential equation is obeyed as for a mechanical
oscillator.)

\bigskip
Summing up the energy over the modes for all frequencies, directions, and polarization
states, one arrives at a zero-point energy density for the electromagnetic fluctuations, and this is
the origin of the electromagnetic quantum vacuum. An energy of $h\nu/2$ per mode of the field
characterizes the fluctuations of the quantized radiation field in quantum field theory. In the
semi-classical representation of SED the quantum vacuum is represented by propagating electromagnetic
plane waves, $\Ezp$ and $\Bzp$, of random phase having this average energy, $h\nu/2$, in each mode.

\bigskip
The volumetric density of modes
between frequencies $\nu$ and
$\nu+d\nu$ is given by the density of states function
$N_{\nu}d\nu=(8\pi\nu^2/c^3)d\nu$. Each state has a minimum $h\nu/2$
of energy,
and using this density of states function and this minimum energy --- that we call the zero-point
energy --- per state one gets the
spectral energy density of the electromagnetic quantum vacuum:
$$\rho(\nu)d\nu={8\pi\nu^2 \over c^3} {h\nu
\over 2} d\nu .
\eqno(1)
$$

\bigskip\noindent
Writing this zero-point radiation together with ordinary blackbody radiation, the energy density is:

$$\rho(\nu,T)d\nu={8\pi\nu^2 \over c^3} \left( {h\nu \over e^{h\nu /kT} -1}
+{h\nu
\over 2}\right) d\nu .
\eqno(2)
$$
The first term (outside the parentheses) represents the mode density, and the
terms inside the parentheses are the average energy per mode of thermal
radiation at temperature $T$ plus the zero-point energy, $h\nu/2$, which has no
temperature dependence. Take away all thermal energy by formally letting $T$
go to zero, and one is still left with the zero-point term. The laws of quantum
mechanics as applied to electromagnetic radiation force the existence of a
background sea of electromagnetic zero-point energy that is traditionally called the electromagnetic
quantum vacuum.

\bigskip
The spectral energy density of eqn. (1) was thought to be no more than a
spatially
uniform constant offset that cancels out when considering energy fluxes, but  it was discovered in
the mid-1970's that the quantum vacuum acquires special
characteristics when viewed from an accelerating frame.
Just as there is an event horizon for a black hole, there is an analogous event
horizon for an accelerating reference frame. Similar to radiation from evaporating black holes
proposed by Hawking [11], Unruh [3] and Davies [4]
determined that a
Planck-like radiation component will arise out of the quantum vacuum in a uniformly-accelerating
coordinate system having constant proper
acceleration {\bf a} (where $|{\bf a}|=a$) with what amounts to an effective
``temperature''

$$T_a = {\hbar a \over 2 \pi c k} .  \eqno(3)$$

\smallskip\noindent
This ``temperature'' characterizing Unruh-Davies radiation does not originate in emission from
particles undergoing thermal
motions.
\footnote{$^a$}{There is likely to be a deep connection
between the fact
that the spectrum that arises in this fashion due to acceleration and
the ordinary
blackbody spectrum have identical form.}
As discussed by Davies, Dray and Manogue [12]:

\medskip
{\parindent 0.4truein \narrower \noindent
One of the most curious properties to be discussed in recent years is the
prediction that
an observer who accelerates in the conventional quantum vacuum of Minkowski
space will
perceive a bath of radiation, while an inertial observer of course
perceives nothing. In
the case of linear acceleration, for which there exists an extensive
literature, the
response of a model particle detector mimics the effect of its being
immersed in a bath of
thermal radiation (the so-called Unruh effect).

}

\bigskip
This ``heat bath'' is a quantum phenomenon. The ``temperature'' is
negligible for most
accelerations. Only in the extremely large gravitational fields of black
holes or in
high-energy particle collisions  can this  become significant. Recently, P. Chen at the Stanford
Linear Accelerator Center has proposed using an ultra high intensity laser to accelerate electrons
violently enough to directly detect Unruh-Davies radiation.[5]

\bigskip
Unruh and Davies treated the electromagnetic quantum vacuum as a scalar field. If a true vectorial
approach is considered there appear additional terms beyond the quasi-thermal Unruh-Davies component.
For the case of no true external thermal radiation
$(T=0)$ but including the acceleration effect
$(T_a)$, eqn. (1) becomes

$$\rho(\nu,T_a)d\nu = {8\pi\nu^2 \over c^3}
\left[ 1 + \left( {a \over 2 \pi c \nu} \right) ^2 \right]
\left[ {h\nu \over 2} + {h\nu \over e^{h\nu/kT_a}-1} \right] d\nu , \eqno(4)$$

\smallskip\noindent where the acceleration-dependent pseudo-Planckian
Unruh-Davies component is placed
after the $h\nu/2$ term to indicate that except for extreme accelerations
(e.g. particle
collisions at high energies) this term is negligibly small.
While these additional acceleration-dependent terms do not show any spatial
asymmetry in
the expression for the spectral energy density, certain asymmetries do
appear when the
momentum flux of this radiation is calculated, resulting in a non-zero Rindler flux.[6]
This asymmetry appears to be the process underlying inertial and gravitational forces.

\bigskip
\centerline{\bf ORIGIN OF THE INERTIAL REACTION FORCE}
\bigskip
Newton's third law states that if an agent applies a force to a point on an object, at
that point there arises an equal and opposite reaction force back upon the agent. In the case of
a fixed object  the equal and opposite reaction force can be traced to
interatomic forces in the neighborhood of the point of contact which act to resist
compression, and these in turn can be traced to electromagnetic interactions
involving orbital electrons of adjacent atoms or molecules, etc.

\bigskip
Now a similar experience of an equal and opposite reaction force arises when a non-fixed object
is forced to accelerate. Why does acceleration create such a reaction force? We suggest that
this equal and opposite reaction force also has an underlying cause which is at least partially
electromagnetic, and specifically may be due to the scattering of electromagnetic quantum vacuum
radiation. RH demonstrated that from the point of view of the pushing agent there exists a net
flux (Poynting vector) of quantum vacuum radiation transiting the accelerating object in a direction
opposite to the acceleration: the Rindler flux.[6] Interaction of this flux with the quarks and
electrons constituting a material object would
create a back reaction force that can be interpreted as inertia.
One
simply needs to assume that there is some dimensionless efficiency factor,
$\eta(\omega)$, that in the case of particles corresponds to whatever the interaction process is (e.g.
dipole scattering). In the case of elementary particles we
suspect that
$\eta(\omega)$ contains one or more resonances --- and in the Appendix discuss why these resonances
likely involve Compton frequencies of relevant particles forming a composite particle or object ---
but this is not a necessary assumption.

\bigskip
The RH approach relies on making transformations of the
$\Ezp$ and $\Bzp$ from a stationary to a uniformly-accelerating coordinate
system (see, for example, $\S$11.10 of Jackson for the relevant transformations [13]). In a stationary
or uniformly-moving frame the
$\Ezp$ and
$\Bzp$ constitute an isotropic radiation pattern. In a uniformly-accelerating frame the
radiation pattern acquires asymmetries. There appears a non-zero Poynting vector in
any accelerating frame, and therefore a non-zero Rindler flux which carries a  net flux of
electromagnetic momentum. The scattering of this momentum flux generates a reaction force,
${\bf F}_r$, proportional to the acceleration. RH found an invariant scalar with the dimension of mass
quantifying the inertial resistance force of opposition per unit of acceleration resulting from
this process. We interpret this scalar as the inertial mass,

$$m_i={V_0 \over c^2} \int \eta(\nu) \rho_{zp}(\nu) \ d\nu , \eqno(5)
$$

\smallskip\noindent
where $\rho_{zp}$ is the well known spectral energy density of the electromagnetic quantum vacuum  of
eqn. (1). In other words, the amount of electromagnetic zero point energy instantaneously transiting
through an object of volume $V_0$ and interacting with the quarks, electrons and all charges in that
object is what constitutes the inertial mass of that object in this view. It is change in the momentum
of the radiation field that creates the resistance to acceleration usually attributed to the inertia
of an object.

\bigskip
Indeed, not only does the ordinary form of Newton's second law, ${\bf F}=m_i{\bf a}$, emerge
from this analysis, but one can also obtain the relativistic form of the second law: [6]

$${\cal F}={d{\cal P} \over d\tau} = {d \over d\tau} (\gamma_{\tau} m_i c, \ {\bf p}
\ ) .
\eqno(6)$$
The origin of inertia, in this picture, becomes remarkably intuitive. Any material object
resists acceleration because the acceleration produces a perceived flux of radiation
in the opposite direction that scatters within the object and thereby pushes against
the accelerating agent. Inertia in the present model appears as a kind of acceleration-dependent
electromagnetic quantum vacuum drag force acting upon electromagnetically-interacting elementary
particles.
The relativistic law for ``mass" transformation involving the Lorentz factor $\gamma$ --- that is, the
formula describing how the {\it inertia} of a body has been calculated to change
according to an observer's relative motion --- is automatically satisfied
in this view, because the correct relativistic form of the reaction force
is derived, as shown in eqn. (6).

\bigskip
\centerline{\bf ORIGIN OF WEIGHT AND THE WEAK EQUIVALENCE PRINCIPLE}
\bigskip
Einstein introduced the {\it local Lorentz invariance} (LLI) principle in order to pass from special
relativity to GR. It is possible to use this principle immediately to extend the results of
the quantum vacuum inertia hypothesis to gravitation (details discussed in two forthcoming papers
[9]).

\bigskip
The idea behind the LLI principle is embodied in the Einstein elevator thought experiment. He
proposed that a freely-falling elevator in a gravitational field is equivalent to one that is not
accelerating and is far from any gravitating body. Physics experiments would yield the same
results in either elevator, and therefore a freely-falling coordinate frame in a gravitational field
is the same as an inertial Lorentz frame. (This is rigorously only true for a ``small
elevator'' since a gravitational field around a planet, say, must be radial, hence there are
inevitably tidal forces which would not be the case for an ideal acceleration.) The device Einstein
used to develop general relativity was to invoke an infinite set of such freely falling frames. In
each such frame, the laws of physics are those of special relativity. The additional features of
general relativity emerge by comparing the properties of measurements made in freely-falling Lorentz
frames ``dropped'' one after the other.

\bigskip
This appraoch of Einstein is both elegant and powerful. The LLI principle immediately tells us that an
object accelerating through the electromagnetic quantum vacuum is equivalent to an object held fixed
in a gravitational field while the electromagnetic quantum vacuum is effectively accelerating
(falling) past it. The prediction of GR that light rays deviate from straight-line propagation in the
presence of a gravitating body --- which Eddington measured in 1919 thereby validating GR ---
translates into acceleration (falling) of the electromagnetic quantum vacuum. An object accelerating
through the electromagnetic quantum vacuum experiences a Rindler flux which causes the inertia
reaction force. A fixed object past which the electromagnetic quantum vacuum is accelerating,
following the laws of GR, experiences the same Rindler flux and the resulting force is what we call
weight. That is why
$m_g=m_i$ and is the basis of the weak equivalence principle.

\bigskip
\centerline{\bf CONCLUSIONS}
\bigskip
Geometrodynamics is an elegant theoretical structure, but there is a very fundamental physics
question that geometrodynamics has never satisfactorily addressed. If an object is forced to deviate
from its natural geodesic motion, a reaction force arises, i.e. the
weight of an object. Where does the reaction force that is weight come from? That same force would
also be the enforcer of geodesic motion for freely falling objects. GR specifies the metric of
spacetime from which geodesics can be calculated, but is there a physical mechanism to keep
freely-falling objects from straying from their proper geodesics? Geometrodynamics does
not provide a physical mechanism for this. It can only claim that deviations of an object from its
proper geodesic motion results in an inertial reaction force. This is true but uninformative. The
quantum vacuum inertia hypothesis provides a physical process generating inertia and weight.

\bigskip
Quantum physics predicts the existence of an underlying sea of zero-point energy at every point in
the universe. This is different from the cosmic microwave background and is also referred to as the
electromagnetic quantum vacuum since it is the lowest state of otherwise empty space. This sea of
energy fills all of space and is absolutely the same everywhere as perceived from a constant velocity
reference frame. But viewed from an accelerating reference frame, the radiation pattern of the energy
becomes minutely distorted: a tiny directional flow is experienced by an accelerating object or
observer, the Rindler flux. Importantly, the force resulting from that energy-momentum flow turns out
to be proportional to the acceleration. When this energy-momentum flow --- that arises automatically
when any object accelerates --- interacts with the fundamental particles constituting matter (quarks
and electrons) a force arises in the direction opposite to the acceleration. This process can be
interpreted as the origin of inertia, i.e. as the basis of Newton's second law of mechanics: ${\bf
F}=m{\bf a}$ (and its relativistic extension).[6]

\bigskip
It has now been discovered that exactly the same distortion of the radiation pattern occurs in
geometrodynamics when the metric is non-Minkowskian.[9] The curved spacetime geodesics of
geometrodynamics affect the zero-point energy in the same way as light rays (because the zero-point
energy is also a mode of electromagnetic radiation).  The gravitational force causing weight and the
reaction force causing inertia originate in an identical interaction with a distortion in the
radiation of the zero-point energy field. Both are a kind of radiation pressure originating in the
electromagnetic quantum vacuum. The underlying distortion of the radiation pattern is due to an event
horizon-like effect and is related to Unruh-Davies radiation and Hawking radiation.

\bigskip
What the quantum vacuum inertia hypothesis accomplishes is to identify the physical process which is
the enforcer of geometrodynamics or general relativity. The quantum vacuum inertia hypothesis
appears to provide a link between light propagation along geodesics and mechanics of material
objects.
Moreover, since the distortion of the zero-point energy
radiation pattern is the same whether due to acceleration or being held stationary in a gravitational
field, this explains a centuries old puzzle: why inertial mass and gravitational mass are the same:
both are due to the same non-zero Rindler flux. This gives us a deeper insight into Einstein's
principle of equivalence.

\bigskip
\centerline{\bf APPENDIX: INERTIA AND THE DE BROGLIE WAVELENGTH}
\bigskip
Four-momentum is defined as
$$
{\bf P}= \left( {E \over c}, \ \ {\bf p} \right) = \left( \gamma m_0 c, \ \ {\bf p} \right)
=\left( \gamma m_0 c, \ \ \gamma m_0{\bf v} \right), \eqno(A1)$$
where $|{\bf P}|=m_0 c$ and $E=\gamma m_0 c^2$. The Einstein-de Broglie relation defines the
Compton frequency
$h \nu_C = m_o c^2$ for an object of rest mass $m_0$, and if we make the de Broglie
assumption that the momentum-wave number relation for light also characterizes matter
then ${\bf p}=\hbar {\bf k}_B$ where ${\bf
k}_B=2\pi(\lambda^{-1}_{B,1},\lambda^{-1}_{B,2},\lambda^{-1}_{B,3})$. We thus write
$$
{{\bf P} \over \hbar} = \left( {2\pi \gamma \nu_C \over c}, {\bf k}_B \right)
= 2 \pi \left( {\gamma \over \lambda_C}, {1 \over \lambda_{B,1}}, {1 \over \lambda_{B,2}}, {1
\over
\lambda_{B,3}} \right)
\eqno(A2)
$$
and from this obtain the relationship
$$
\lambda_B={c \over \gamma v} \lambda_C \eqno(A3)
$$
between the Compton wavelength, $\lambda_C$, and the de Broglie wavelength, $\lambda_B$. For a
stationary object $\lambda_B$ is infinite, and the de Broglie wavelength decreases in inverse
proportion to the momentum.

\bigskip
Eqn. (5) is very suggestive that quantum vacuum-elementary particle interaction involves
a resonance at the Compton frequency. 
De Broglie proposed that an elementary particle is associated with a localized
wave whose frequency is the Compton frequency.
As summarized by Hunter [14]: ``\dots what we regard as the (inertial) mass of the particle
is, according to de Broglie's proposal, simply the vibrational energy (divided by $c^2$)
of a localized oscillating field (most likely the electromagnetic field). From this
standpoint inertial mass is not an elementary property of a particle, but rather a
property derived from the localized oscillation of the (electromagnetic) field. De Broglie
described this equivalence between mass and the energy of oscillational motion\dots as
{\it `une grande loi de la Nature'} (a great law of nature).'' 

\bigskip
This perspective is
consistent with the proposition that inertial mass, $m_i$, may be a coupling parameter between
electromagnetically interacting particles and the quantum vacuum. Although De Broglie assumed
that his wave at the Compton frequency originates in the particle itself (due to some
intrinsic oscillation or circulation of charge perhaps) there is an alternative interpretation
discussed in some detail by de la Pe\~na and Cetto that a particle  ``is tuned to a wave
originating in the high-frequency modes of the zero-point background field.''[8] The de Broglie
oscillation would thus be due to a resonant interaction with the quantum vacuum, presumably the same
resonance that is responsible for creating a contribution to inertial mass as in eqn. (5). In other
words, the electromagnetic quantum vacuum would be driving this
$\nu_C$ oscillation.

\bigskip
We therefore suggest that an elementary charge driven to oscillate at the Compton
frequency, $\nu_C$,  by the quantum vacuum may be the physical basis of the $\eta(\nu)$
scattering parameter in eqn. (5).  For the case of the electron, this would imply that
$\eta(\nu)$ is a sharply-peaked resonance at the frequency, expressed in terms of
energy, $h\nu_C=512$ keV. The inertial mass of the electron would physically be the reaction
force due to resonance scattering of the electromagnetic quantum vacuum radiation, the Rindler flux,
at that frequency.

\bigskip
{\it This leads to a surprising corollary.} It has been shown that as viewed from a
laboratory frame, a standing wave at the Compton frequency in the electron frame transforms
into a traveling wave having the de Broglie wavelength
for a moving electron.[8,14,15,16] The wave nature of the moving electron (as measured in the
Davisson-Germer experiment, for example) would be basically due to Doppler shifts associated with
its Einstein-de Broglie resonance at the Compton frequency.
A simplified heuristic model shows this, and a detailed treatment showing the same result
may be found in de la Pe\~na and Cetto [8]. Represent a quantum vacuum-like driving force field as two
waves having the Compton frequency $\omega_C=2\pi \nu_C$ travelling in equal and opposite
directions,
$\pm
\hat{x}$. The amplitude of the combined oppositely-moving waves acting upon an electron will be
$$
\phi=\phi_+ + \phi_{-}=2 \cos \omega_C t \cos k_C x. \eqno(A4)
$$

\bigskip
But now assume an electron is moving with velocity $v$ in the $+x$-direction. The wave
responsible for driving the resonant oscillation impinging on the electron from the front
will be the wave seen in the laboratory frame to have frequency $\omega_-=\gamma
\omega_C (1 - v/c)$, i.e. it is the wave below the Compton frequency in the laboratory
that for the electron is Doppler shifted up to the
$\omega_C$ resonance. Similarly the zero-point wave responsible for driving the electron resonant
oscillation impinging on the electron from the rear will have a laboratory frequency
$\omega_+=\gamma \omega_C (1 + v/c)$ which is Doppler shifted down to $\omega_C$ for the
electron. The same transformations apply to the wave numbers,
$k_+$ and $k_-$. The Lorentz invariance of the electromagnetic quantum vacuum spectrum ensures that
regardless of the electron's (unaccelerated) motion the up- and down-shifting of the laboratory-frame
spectral energy density will always yield a standing wave in the electron's frame.

\bigskip
It can be shown [8,15] that the superposition of these two oppositely-moving, Doppler-shifted waves is
$$
\phi'=\phi'_++\phi'_{-}=2 \cos(\gamma \omega_C t - k_B x) \cos(\omega_B t - \gamma k_C x).
\eqno(A5)
$$
Observe that for fixed $x$, the rapidly
oscillating ``carrier'' of frequency $\gamma \omega_C$ is modulated by the slowly varying
envelope function in frequency $\omega_B$. And {\it vice versa} observe that at a given $t$ the
``carrier'' in space appears to have a relatively large wave number $\gamma k_C$ which is
modulated by the envelope of much smaller wave number $k_B$. Hence
both timewise at a fixed point in space and spacewise at a given time, there appears a
carrier that is modulated by a much broader wave of dimension corresponding to the de
Broglie time $t_B=2\pi/\omega_B$, or equivalently, the de Broglie wavelength
$\lambda_B=2\pi/k_B$.

\bigskip
This result may be generalized to include quantum vacuum radiation from all other directions, as may
be found in the monograph of de la Pe\~na and Cetto [8]. They conclude by stating:
``The foregoing discussion assigns a physical meaning to de Broglie's wave: it is the {\it
modulation} of the wave formed by the Lorentz-transformed, Doppler-shifted superposition
of the whole set of random stationary electromagnetic waves of frequency
$\omega_C$ with which the electron interacts selectively.''

\bigskip
Another way of looking at the spatial modulation is in terms of the
wave function: the  spatial modulation of eqn. (A5) is exactly the $e^{i p x / \hbar}$ wave
function of a freely moving particle satisfying the Schr\"odinger equation.
The same argument has been made by Hunter [14].
In such a view the quantum wave function
of a moving free particle becomes a ``beat frequency'' produced by the relative motion of
the observer with respect to the particle and its oscillating charge.

\bigskip
It thus appears that a simple model of a particle as an electromagentic quantum vacuum-driven
oscillating charge with a resonance at its Compton frequency may simultaneously offer insight into
the nature of inertial mass, i.e. into rest inertial mass and its relativistic extension, the
Einstein-de Broglie formula and into its associated wave function involving the de Broglie
wavelength of a moving particle. If the de Broglie oscillation is indeed driven by the
electromagnetic quantum vacuum, then it is a form of Schr\"odinger's {\it zitterbewegung}. Moreover
there is a substantial literature attempting to associate spin with {\it zitterbewegung} tracing back
to the work of Schr\"odinger [17]; see for example Huang [18] and Barut and Zanghi [19].

\bigskip
{

\bigskip

\parskip=0pt plus 2pt minus 1pt\leftskip=0.25in\parindent=-.25in 

\centerline{\bf REFERENCES}
\bigskip

1] M. Jammer, Concepts of Mass in Contemporary Physics and Philosopy,
Princeton Univ. Press (2000)

[2] B. Haisch, A. Rueda, and H. E. Puthoff, Phys. Rev. A {\bf 49} 678
(HRP) (1994) 

[3] W. G. Unruh, Phys. Rev. D, {\bf 14} 870 (1976)

[4] P. C. W. Davies, J. Phys. A, {\bf 8} 609 (1975)

[5] P. Chen, reported at American Astronomical Society meeting, June 6,
2000.

[6] A. Rueda and B. Haisch, Found. Phys. {\bf 28}  1057
(1998); A. Rueda and B. Haisch, Phys. Lett. A {\bf 240} 115  (1998)

[7] T. H. Boyer, Phys. Rev. D, {\bf 11} 790 (1975) 

[8] L. de la Pe\~na, and A. M. Cetto, The Quantum Vacuum: An
Introduction to Stochastic Electrodynamics, Kluwer Acad. Publ. (1996)

[9] A. Rueda, B. Haisch and R. Tung, in prep. (2001); R. Tung, B. Haisch
and A. Rueda, in prep. (2001)

[10] R. Loudon, The Quantum Theory of Light (2nd ed.) Clarendon Press,
Oxford (1983)

[11] S. Hawking, Nature, {\bf 248} 30 (1974)

[12] P. C. W. Davies, T. Dray, and C. A. Manogue, Phys. Rev. D {\bf 53}
4382 (1996)

[13] J. D. Jackson, Classical Electrodynamics (3rd ed.) (1999)

[14] G. Hunter, in The Present Status of the Quantum Theory of Light, S.
Jeffers et al. (eds.), (Kluwer Acad. Publ.), chap. 12 (1996)

[15] A. F. Kracklauer, Physics Essays {\bf 5} 226 (1992); for a formal
derivation and further illuminating discussion see Chap. 12 of [4]

[16] B. Haisch and A. Rueda, Phys. Lett. A, {\bf 268}, 224 (2000) 

[17] E. Schr\"odinger, Sitz. Ber. Preuss. Akad. Wiss. Phys.-Math. Kl, {\bf
24}, 4318 (1930) 

[18] K. Huang, Am. J. Phys. {\bf 20} 479 (1952) 

[19] A. O. Barut and N. Zanghi, Phys. Rev. Lett. {\bf 52} 209 (1984) 

}

\bye